\documentclass[prints]{aa}
\usepackage{psfig}
\begin{document}
\thesaurus{11.04.2; Galaxies: dwarf
	   11.09.1; NGC 6946 Group
	   11.16.1; Galaxies: photometry
}

\title{A group of galaxies around the giant spiral NGC~6946}
\titlerunning{A group of galaxies around NGC~6946}

\author{I.D.Karachentsev \inst{1} \and M.E.Sharina \inst{1}
\and W.K.Huchtmeier \inst{2}}

\institute{Special Astrophysical Observatory, Russian Academy of Sciences,
      N.Arkhyz, KChR, 369167, Russia
\and Max-Plank-Institut f\"{u}r Radioastronomie, Auf dem H\"{u}gel 69, 53121,
      Bonn, Germany}
\date{Received 15 May 2000, Accepted . . .}
\maketitle
\begin{abstract}

  We present large-scale CCD images of a dozen low surface brightness
objects in the vicinity of giant Sc galaxy NGC~6946, which have been found
recently on POSS-II copies. Six of them: UGC~11583, kk251, kk252, kkr55,
kkr56, and kkr59 are dwarf irregular galaxies resolved into stars. We
determined their distances from the luminosity of the brightest blue stars.
The mean estimated distance to the NGC~6946 group is 5.9$\pm$0.4 Mpc.
Together with Cepheus 1, discovered before, the group consist of 8 late
type galaxies. Almost all of them are detected or marginally detected
in the HI line. The group is characterized by a linear projected
diameter of 420 kpc, a radial velocity dispersion of 95 km~s$^{-1}$, and a
virial mass-to-luminosity ratio about 56 $M_{\sun}/L_{\sun}$.
\end{abstract}

\section{Introduction}

  NGC 6946 is amongst the seven more massive galaxies in the Local
Volume (=LV) with distances  $D <$ 7 Mpc (Karachentsev et al. 1999).
51\% of the population of the LV are within the gravitational influence of
these "oligarchs". Each of 7 the
giant galaxies, except NGC~6946, has a lot of dwarf companions: from 7--9
members around Sc galaxies M101 and NGC~5236 to 32 members around giant
elliptical galaxy NGC~5128. In this sense the absence of accompying dwarf
systems around NGC~6946 within a radius of $\sim$1 Mpc (Tully, 1988) looked
like a remarkable anomaly. However, the apparent isolation of NGC~6946 may be
related with its location on a border of the Local Void (Tully 1988).
The substantial galactic extinction in the direction of NGC~6946 has also
an influence on the visibility of faint dwarf galaxies.

  Based on copies of the Second Palomar Observatory Sky Survey (=POSS-II),
Karachentseva \& Karachentsev (1998) undertook a search for galaxies of low
surface brightness in the vicinity of normal galaxies in the LV. Observations
of these galaxies (kk- list) with the 100-m Effelsberg radio telescope
revealed three dwarf galaxies: UGC~11583 = kk250, kk251, and kk252, with
positions and radial velocities close to NGC~6946 (Huchtmeier et al. 1997).
Using the luminosity of the brightest stars in NGC~6946 and UGC~11583, Sharina
et al. (1997) derived distance moduli to these galaxies: 28.90 and 29.13 mag,
respectively. During the course of a search for compact high velocity clouds
Burton et al. (1999) discovered beside NGC~6946 a low surface brightness
galaxy with radial velocity $V_h= +58$ km~s$^{-1}$ named by them Cepheus 1. Subsequent
searches for nearby dwarf galaxies by Karachentseva et al. (1999) (kkr- list)
and Karachentseva et al. (2000) (kkh- list) led to the discovery of some 
more LSB
objects around NGC~6946. In this paper we present results of optical and
HI observations of these new objects, particularly data on three new probable
companions of NGC~6946.

\section{Optical observations and photometry}

Table 1 lists 18 nearby dwarf galaxy candidates from the above mentioned
list with projected distances to NGC~6946 of less than 1 Mpc. For the
giant galaxy NGC~6946, also included in the Table, a distance of 5.9 Mpc
(see Sect.4) is adopted. Besides the galaxy name, and its equatorial and
galactic coordinates, Table~1 contains: major and minor axes
measured on the blue (J) POSS-II copy, morphological type determined with
with POSS-II copies, surface brightness (low, very low, extremely low), and
galactic extinction in the B, and I bands according to Schlegel et al.(1998).

  The HI line observations showed that objects kkr46, kkr48, and kkr58
are background galaxies with heliocentric radial velocities of respectively
+3779, +3158, and +2756 km~s$^{-1}$. Images of the remaining 15 objects
from the Digital Sky Survey are reproduced in Figure~1, each
field being 5 arcmin wide .

  Most objects were imaged with the 6-m telescope (Russia) and the
2.5-m Nordic telescope (Spain) in different filters of the Johnson-Cousins
system. Table~2 presents the observational log with indication of exposure
time and seeing. A 1k$\times$1k CCD detector was used at the 6-m telescope,
providing a total view of 3$\farcm5$ with a resolution of 0$\farcs21$/pixel.

\begin{figure*}[hbt]

\caption{Digital Sky Survey images of 15 dwarf galaxy candidates in the
vicinity
       of NGC~6946. The field size is 5$\arcmin$, North is to the top, and East is to the left.}
\end{figure*}

\begin{table*}[hbt]
\caption{List of dwarf galaxy candidates around NGC 6946
	  within projected radius of $\sim$1 Mpc.}
\begin{tabular}{lccrrrrllrc} \\ \hline
\multicolumn{1}{c}{Name}&
\multicolumn{2}{c}{RA (2000.0) DEC}&
\multicolumn{2}{c}{$L, B$}&
\multicolumn{2}{c}{$a\times b$}&
\multicolumn{2}{c}{Type}&
\multicolumn{1}{c}{$A_B$}&
\multicolumn{1}{c}{$A_I$} \\ \hline

kkr46 &  19$^h$36$^m$36$\fs$2&  54$\degr$38$\arcmin$21$\arcsec$ & 86.8$\degr$& 15.7$\degr$&  0$\farcm$8&  0$\farcm$6&   Ir&  L &  0.456&  0.205\\
kkh90 &  19 41 50.6&  68 34 14& 100.5& 20.7&  1.3&  0.75&  Ir? & VL & 0.673&  0.303 \\
kkr48 &  19 57 56.4&  62 37 21&  95.5& 16.7&  0.9&  0.45&  Ir  & L  & 0.359&  0.162 \\
kkh92 &  20 10 01.1&  66 05 01&  99.4& 17.2&  0.5&  0.2 &  Ir  & L  & 1.414&  0.636 \\
kkr51 &  20 21 15.5&  52 28 03&  88.2&  8.9&  0.7&  0.35&  Ir  & L  & 1.152&  0.518 \\
      &            &          &      &     &     &      &      &    &      &        \\
kk250 &  20 30 15.0&  60 26 31&  95.6& 12.3&  1.8&  0.8 &  Ir  & VL & 1.319&  0.593 \\
kk251 &  20 30 32.9&  60 21 13&  95.6& 12.2&  1.6&  0.8 &  Ir? & VL & 1.236&  0.556 \\
kk252 &  20 31 33.1&  60 48 48&  96.0& 12.4&  0.9&  0.9 &  Sph?& VL & 1.913&  0.860 \\
kk254 &  20 34 45.7&  61 05 37&  96.5& 12.2&  1.5&  0.9 &  Ir? & EL & 1.936&  0.870 \\
N6946 &  20 34 51.9&  60 09 15&  95.7& 11.7& 11.0& 10.0 &  Sc  & -  & 1.475&  0.663 \\
      &            &          &      &     &     &      &      &    &      &        \\
kkr55 &  20 45 20.8&  60 24 40&  96.7& 10.8&  0.6&  0.4 &  Ir  & L  & 2.941&  1.322 \\
kkr57 &  20 47 32.6&  63 04 10&  99.0& 12.2&  0.5&  0.45&  Ir  & L  & 1.444&  0.649 \\
kkr56 &  20 48 24.1&  58 37 06&  95.5&  9.4&  0.7&  0.45&  Ir  & L  & 3.135&  1.409 \\
kkr58 &  20 49 33.7&  58 06 18&  95.2&  8.9&  2.1&  0.2 &  Sm  & VL & 3.635&  1.634 \\
Ceph1 &  20 51 10.7&  56 53 25&  94.4&  8.0&  3.0:&  1.5:&  Sm?& VL & 4.047&  1.819 \\
      &            &          &      &     &     &      &      &    &      &        \\
kkh93 &  20 57 59.5&  62 20 56&  99.2& 10.8&  0.5&  0.35&  Ir  & L  & 2.891&  1.300 \\
kkr59 &  21 03 24.2&  57 17 14&  95.8&  7.0&  2.3&  1.4 &  Ir  & VL & 3.863&  1.737 \\
kkr60 &  21 05 53.0&  57 12 19&  95.9&  6.7&  0.7&  0.5 &  Ir  & VL & 4.574&  2.056 \\
kkr62 &  21 30 44.9&  52 41 39&  95.2&  1.0&  1.1:& 0.6:&  Ir? & EL &12.062&  5.423  \\
\hline
\end{tabular}
\end{table*}
\begin{table*}[hbt]
\caption{Observational log}
\scriptsize
\begin{tabular}{lcccccl}\\
\hline
 Object&    Date   & Telescope&  Filter&  Exposure&  Seeing &  Comments \\ \hline
       &           &          &        &          &         &           \\
 kkh90 &  18.06.99 &  BTA     &   V    &   600s   &   1.0   &  cirrus   \\
       &           &          &   I    &   600    &         &           \\
 kkh92 &  10.07.99 &  BTA     &   R    &   300    &   1.1   &  distant  \\
       &  06.11.99 &  BTA     &   R    &   600    &   2.0   &           \\
       &           &          &   I    &   600    &         &           \\
 kk250=&  10.07.99 &  BTA     &   R    &   300    &   1.1   &  resolved \\
 U11583&           &          &        &          &         &           \\
 kk251 &  26.07.97 &  NOT     &   I    &   600    &   0.6   &  resolved \\
       &  10.07.99 &  BTA     &   R    &   300    &   1.1   &           \\
       &  10.07.99 &          &   I    &   300    &         &           \\
 kk252 &  28.07.97 &  NOT     &   V    &   900    &   0.7   &  resolved \\
       &           &          &   I    &   600    &         &           \\
 kk254 &  26.07.97 &  NOT     &   I    &   600    &   0.7   &  cirrus   \\
       &  10.07.99 &  BTA     &   R    &   300    &   1.1   &           \\
       &           &          &   I    &   300    &         &           \\
 kkr55 &  09.07.99 &  BTA     &   V    &   600    &   1.2   &  resolved \\
       &           &          &   R    &   300    &         &           \\
       &           &          &   I    &   300    &         &           \\
 kkr57 &  09.07.99 &  BTA     &   V    &   600    &   1.2   &  semi-resolved\\
       &           &          &   R    &   300    &         &           \\
 kkr56 &  09.07.99 &  BTA     &   R    &   300    &   1.2   &  resolved \\
       &           &          &   I    &   300    &         &           \\
 kkr58 &  10.07.99 &  BTA     &   R    &   300    &   1.1   &  distant edge-on\\
       &           &          &        &          &         &  RFGC galaxy \\
 kkh93 &  10.07.99 &  BTA     &   R    &   300    &   1.1   &  distant  \\
       &  06.11.99 &  BTA     &   R    &   600    &   2.0   &           \\
       &           &          &   I    &   600    &         &           \\
 kkr59 &  10.07.99 &  BTA     &   R    &   300    &   1.1   &  resolved \\
       &           &          &   I    &   300    &         &           \\
 kkr62=&  06.11.99 &  BTA     &   R    &   900    &   2.5   &  PN/RN ?  \\
 Dw95+1&           &          &   I    &   900    &         &           \\
\hline
\end{tabular}
\end{table*}
\begin{figure*}
\caption{CCD images of twelve LSB objects around NGC~6946: a) kkh90, b) kkh92,
       c) kk250= UGC~11583, d) kk251, e) kk252, f) kk254, g) kkr55, h) kkr57, i) kkr56,
       j) kkh93, k) kkr59, and l) kkr62. All of them, except kk252, were
       observed with the 6-m telescope. On each frame a scale is indicated
       with a 10$\arcsec$ bar.}
\end{figure*}

The observations at the Nordic telescope were performed with a 2k$\times$2k CCD
providing a field of 3$\farcm7$ with a resolution of 0$\farcs22$/pixel after
2$\times$2 rebinning.
  The images were processed with the DAOPHOT II program (Stetson
et al. 1998) implemented in MIDAS. Standard stars from Landolt (1992) were
used for calibration. In cases where galaxies are resolved into stars we
derived colour-magnitude diagrams for stars in the galaxy and in
surrounding field. Besides stellar photometry we also carried out aperture
photometry of the galaxies in circular apertures that allows the total
integral magnitudes and surface brightness profiles to be measured. The 
luminosity of the brightest stars was used to determine
distances. Below we consider each object in more detail.

{\em kkh90.} The R- band image of this object is shown in Fig. 2a. The object
has a hook shape and is unresolved into stars to the limiting magnitude
R$\sim$24 mag. It is undetected in the HI line. kkh90 is very probably
an isolated fragment of a faint reflexion nebula.

{\em kkh92.} In Fig. 2b kkh92 looks like a distant
galaxy, which has an elongated central part (bar?) surrounded with a LSB
envelope (disk?). Its HI spectrum shows no emission within
[$-$500, +4000] km~s$^{-1}$.

{\em UGC11583 = kk250.} This irregular dwarf galaxy 
has been discussed by Sharina et al. (1997) as a companion of NGC~6946. 
Fig. 2c presents its R image derived with a
seeing of FWHM = 1$\farcs$1. According to Schlegel et al. (1998)
the galactic extinction in the direction of UGC~11583, $A_B$ = 1.32 mag, is
lower than adopted before (1.93 mag), increasing its
distance modulus.

{\em kk251.} Its angular separation from the previous galaxy is only 6$\arcmin$,
less than the half power width (9$\farcm$3) of the 100-m radio telescope.
The similarity of their radial velocities, +127 and +126 km~s$^{-1}$
may be the result of HI flux confusion. However, Fig. 2d suggests that
kk251 is a nearby irregular galaxy resolved into stars. Its distance
estimate (see below) is in agreement with membership in the NGC~6946
group. The integral magnitude of kk251, $B_T$ = 16.49, has been measured by
Hopp (1998).

{\em kk252.} A rather bright red star is projected into the western side
of the galaxy (Fig. 2e). Judging on its radial velocity, $V_h = +$132 km~s$^{-1}$,
this well- resolved irregular galaxy is a companion of NGC~6946.

{\em kk254.} The apparent structure of this object (Fig. 2f) definitely shows
that it is a small reflexion nebula. Such isolated cirrus with very low
surface brightness can easily be confused with an irregular galaxy.
The object is undetected in HI.

{\em kkr55.} This compact irregular galaxy is well resolved into
stars (Fig. 2g). The brightest stars beside its center are embedded
into a diffuse envelope suggesting a starburst region. The galaxy is
undetected in the HI line, but its HI emission is probably merged with
the strong emission of the local galactic hydrogen (see below).

{\em kkr57.} The central part of the galaxy has an asymmetric, skewed shape
(Fig. 2h). Apparently kkr57 is not resolved into stars but into compact knots.
It is unclear why such a blue irregular galaxy is undetected in HI.
Spectral observations of the object in the H alpha line may clarify its
nature.

{\em kkr56.} As it is seen from Fig. 2i, this irregular galaxy is well resolved
into stars. In spite of its marginal detection in the HI line, kkr56 seems
to be a true companion of NGC~6946. This should be confirmed by
observations in H alpha.

{\em kkh93.} This galaxy is undetected in HI. Probably, it is a medium distant
Sm galaxy (Fig. 2j).

{\em kkr59.} On POSS-II plates this irregular galaxy looks like
a planetary nebula. Its large-scale image, Fig. 2k, reveals a lot of faint
stars, as well as several probable compact HII regions on the northern side
of the galaxy. A marginal detection of kkr59 in the HI line needs to be
confirmed by optical spectroscopy.

{\em kkr62.} This object was imaged under bad seeing (Fig. 2l). It is located
in a zone of very high extinction ($A_B\sim12$ mag!) near the HI- source
Dwingeloo 095+1.0 with a velocity of $V_h= +$159 km~s$^{-1}$ (Henning et al. 1998).
kkr62 has an extremely low surface brightness in the R band and is practically
unseen in the I band. It might be a planetary nebula or a small
reflexion nebula.

\section{HI observations}

  All the objects in Table 1 were observed in the HI 21 cm line using the
100-m radio telescope at Effelsberg (HPBW= 9$\farcm$3). Observations have been
obtained in the total power mode combining the on-source position with a 
reference
field. A velocity resolution of about 5 km~s$^{-1}$ was achieved after Hanning
smoothing. The search range was [$-$500, +4000] km~s$^{-1}$. As mentioned above,
three dwarf irregular galaxy: kk250, kk251, and kk252 have been detected
before (Huchtmeier et al. 1997). For three other dwarfs: kkr55, kkr56,
and kkr59, the HI line profiles are presented in Fig. 3.
\begin{figure}[hbt]
 \vbox{\includegraphics{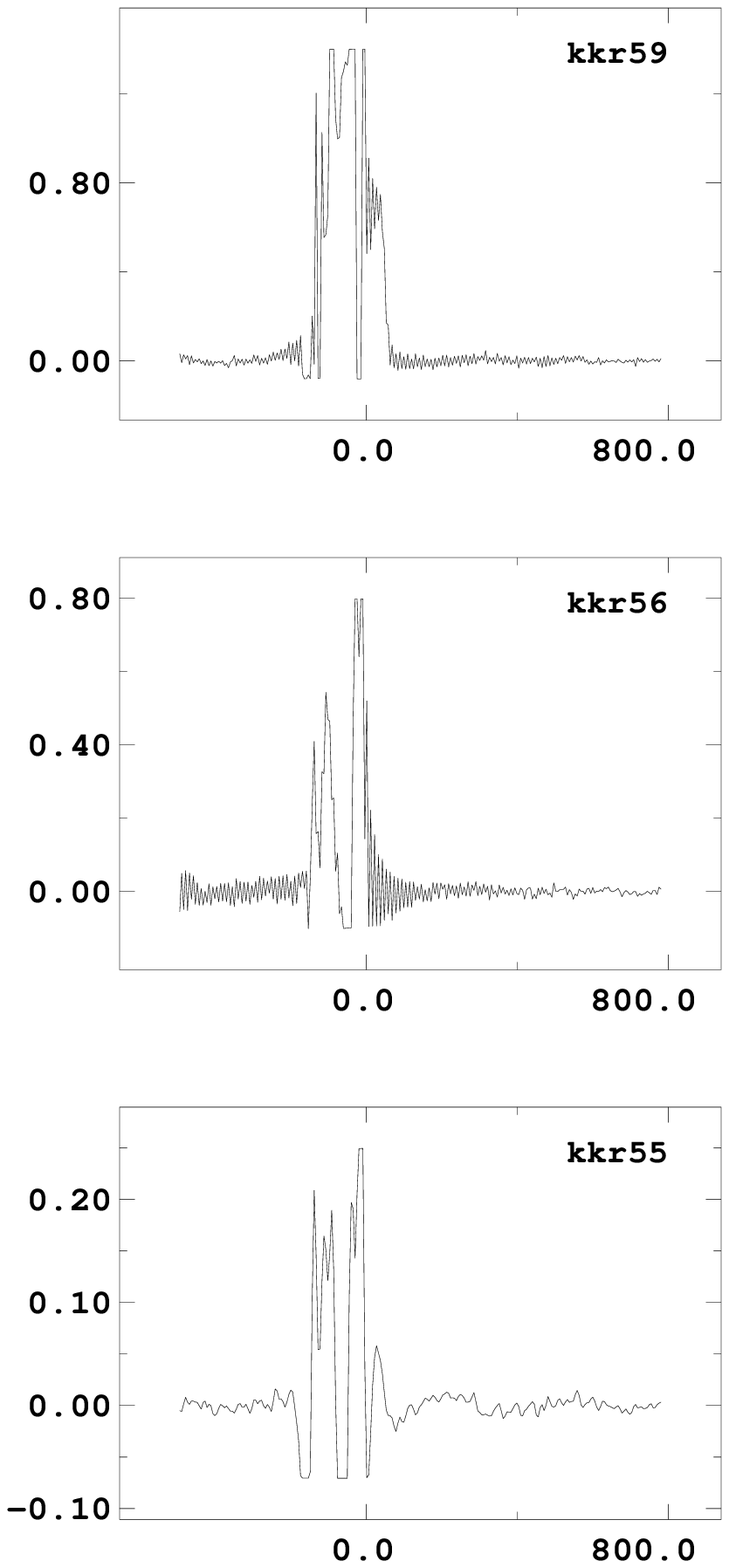}}\par
\vspace{20cm}
\caption{HI spectra of the galaxies: a) kkr55, b) kkr56, and c) kkr59, obtained
       with the 100-m Effelsberg radio telescope.}
\end{figure}

  The main difficulty of detection in the HI line of galaxies around
NGC6946 is caused by their low radial velocities which are expected to be
in the range of [$-$150, +250] km~s$^{-1}$. As seen from Fig. 3 as well as
Fig. 1 of Burton et al. (1999), the galactic hydrogen emission is
predominant in the velocity range of about [$-$100, 0] km~s$^{-1}$. It 'hides'
almost completely weak emission from galaxies with velocities in the
indicated range, and leads to an asymmetric distribution of the group
members according to their velocities. We find a similar situation 
in two other nearby groups, M81 and IC342/Maffei (van Driel et al.
1998, Huchtmeier et al. 2000).

  In the case of kkr55 the situation seems to be very complicated. A weak
emission feature with $V_h$ = 23 km~s$^{-1}$, a line width 
$W_{50}\sim$44 km~s$^{-1}$, and a flux
$F\sim$1 Jy km~s$^{-1}$ is suspected. Radio observations with an 
aperture synthesis
telescope or optical spectra may help to verify this feature. In the HI line
spectrum of kkr56 we note possible weak
\begin{table}[hbt]
\caption{Apparent magnitudes of the observed galaxies}
\begin{tabular}{ccccr}\\ \hline
 Galaxy  &     $I_T$ &    $R_T-I_T$  &   $I_c$  &     $h\arcsec$ \\
	 &         &   $[V - I]_T$ &        &        \\               \hline
	 &         &             &        &        \\
 kkh92   &    15.58&     0.62    &   22.1 &     9  \\
 kk251   &    14.42&     0.78    &   22.8 &    21  \\
 kk252   &    14.16&    [1.41]   &   21.0 &     7  \\
 kkr55   &    14.57&     0.73    &   21.5 &    10  \\
 kkr56   &    14.69&     0.91    &   22.5 &    11  \\
 kkh93   &    15.41&     0.72    &   21.4 &     5  \\
 kkr59   &    12.52&     0.98    &   22.5 &    40  \\
\hline
\end{tabular}
\end{table}
\begin{table*}[hbt]
\caption{Observed properties of galaxies in the NGC~6946 group}
\begin{tabular}{lcccccccc} \\ \hline

 Parameter&   N6946&   kk250&   kk251&   kk252&   kkr55&   kkr56&   Ceph1&   kkr59\\ \hline
 Type     &    Sc  &    Ir  &    Ir  &    Ir  &    Ir  &    Ir  &    Sm  &    Im  \\
 $B_T$      &   9.7  &  15.7  &  16.5  &  16.7  &  17.0  &  17.6  &  15.4  &  15.7  \\
 $A_B$      &   1.48 &   1.32 &   1.24 &   1.91 &   2.94 &   3.14 &   4.05 &   3.86 \\
$(B-V)_{T,0}$    & 0.55   &    ---   &   0.79:&   0.47 &   0.29 &   0.44 &   0.3: &   0.38 \\
$<B(3B)>_{3,0}$   &  19.34 &  21.70 &  21.37 &  21.22 &  20.96 &  21.37 &    ---   &  19.98 \\
$<B-V>_{3,0}$    &  $-$0.02 &   0.30 &   0.07 &  $-$0.21 &   0.27 &   0.13 &    ---   &  $-$0.28 \\
$(m-M)_0$    &  29.15 &  29.57 &  28.63 &  28.64 &  28.66 &  29.04 & [28.89]&  28.34 \\
 $D$, Mpc    &   6.8  &   8.2  &   5.3  &   5.3  &   5.4  &   6.4  &  [6.0] &   4.7  \\
$M_{B,0}$      &  $-$20.6 &  $-$14.5 &  $-$13.6 &  $-$14.1 &  $-$14.8 &  $-$14.4 &  $-$17.5 &  $-$17.0 \\
 $V_h$, km~s$^{-1}$ &  +51   & +127   & +126   & +132   &  +23?  & $-$135:  &  +58   &  +17:  \\
 $V_o$, km~s$^{-1}$ &  +343  & +430   & +429   & +435   &  +330? & +172:  &  +374  &  +327: \\
 $W_{50}$, km~s$^{-1}$&  169   &   90   &   64   &   27   &    44? &   10:  &    90  &    63: \\
 $F$, Jykm~s$^{-1}$ &  839   &   20   &   14.6:&   1.36 &    1?  &   5.6: &   136  &    36: \\
$M(HI)/L_B$  &  0.26  &  1.86  &  3.02: &   0.18 &   0.07?&   0.56 &   0.76 &   0.32 \\
 $r, \degr$  &   0    &  0.64  &  0.57  &   0.78 &   1.35 &   2.31 &   3.90 &   4.68 \\
 $R$, kpc   &   0    &   66   &   59   &   81   &   139  &   239  &   403  &   483  \\ \hline
\end{tabular}
\end{table*}
emission with parameters: $V_h =-135$ km~s$^{-1}$, $W_{50}$ = 10 km~s$^{-1}$, and $F$ = 5.6 Jy km~s$^{-1}$.
The third galaxy, kkr59, has
probably emission with $V_h = +$17, $W_{50}$ = 63, and $F$ = 36 Jy km~s$^{-1}$ 
adjacent
to the Local HI emission. All three HI
detections are marginal and have to be confirmed with independent
optical/radio observations.

\section{Photometry results and distances}

  Results of our surface photometry of 7 galaxies around NGC~6946 are
presented in Table~3. Its columns contain: (1) galaxy name, (2) integral
apparent magnitude in the $I$ band, (3) integral colour $R-I$ or $V-I$, (4)
central surface brightness in the $I$ band, (5) exponential scale lenght
in arcsec. To estimate the integral blue magnitude of a galaxy we used the
transformation  $(B-V)_0 = 0.85\cdot(V-I)_0 - 0.20$,  derived by Makarova (1999)
for late type galaxies. When only the integral colour $R-I$ is measured,
we used an approximate relation $(R-I)_0 = 0.48(V-I)_0$
after dereddening.

  Based on photometry of the brightest stars in the galaxies: kk251, kk252,
kkr55, kkr56, and kkr59 we determine the mean apparent magnitude of the
three
brightest blue stars, $<B(3B)>$, and use it as an indicator of distance to
the galaxies. According to Sandage \& Tammann (1974) and de Vaucouleurs
(1978) the mean luminosity of blue supergiants in a late type galaxy
depends on its integral magnitude $B_T$. After calibration with cepheids, this
relation can be written (Karachentsev \& Tikhonov, 1994) as
    $$(m-M)_0 = <B(3B)> - A_b +0.51[<B(3B)> - B_T] + 4.14.$$
In order to obtain B magnitudes from V or R we used the empirical relations:
$$    (R-I)_0 = 0.48(V-I)_0,$$
$$    (B-V)_0 = 0.89(V-I)_0  ,$$
determined via standard Landolt's stars with colours $(B-V)_0 < 1$. The
derived distance moduli are presented in Table~4, which
contains the basic integral parameters of the galaxies. In this Table we
incuded also data on Cepheus 1 from Burton et al. (1999), as well as data on
NGC6946 and kk250 from Sharina et al. (1997)  with the corrections for
extinction from Schlegel et al. (1998). In this Table the rows indicate the
following parameter: 
(1) morphological type, (2) integral galaxy magnitude, (3) adopted value
of extinction; (4) integral galaxy colour corrected for extinction; (5,6)
the mean apparent magnitude and the mean colour of the three brightest
blue stars, corected for redening, (7,8) derived distance modulus and linear
distance in Mpc; (9) absolute magnitude assuming  the mean modulus
of $<m-M>_0 = 28.86$; (10,11) heliocentric velocity and velocity reduced to
the Local Group centroid; (12) the HI line width at a 50\% level of the
maximum; (13) the HI line flux; (14) hydrogen mass-to-luminosity ratio in
solar units; (15,16) angular and linear projected separation from NGC~6946.

\section{Properties of the NGC~6946 group}

In spite of substantial extinction towards the NGC~6946 group and of the indirect
estimate of blue magnitudes of member stars, the derived distance moduli of
the candidate group members
agree satisfactorily. Assuming a distance modulus of 28.89 for
Cepheus 1 (Burton et al. 1999) from the luminosity of its HII regions,
the mean distance modulus for eight members of the group is 28.86$\pm$0.14
with a standard deviation of 0.36 mag, typical for distances using the brightest
stars. With a mean distance of 5.9$\pm$0.4 Mpc, the absolute magnitudes of
dwarf irregular companions of NGC~6946 lie in a usual range of [$-$13.6,
$-$17.5].
\begin{figure}[hbt]
 \vbox{\includegraphics{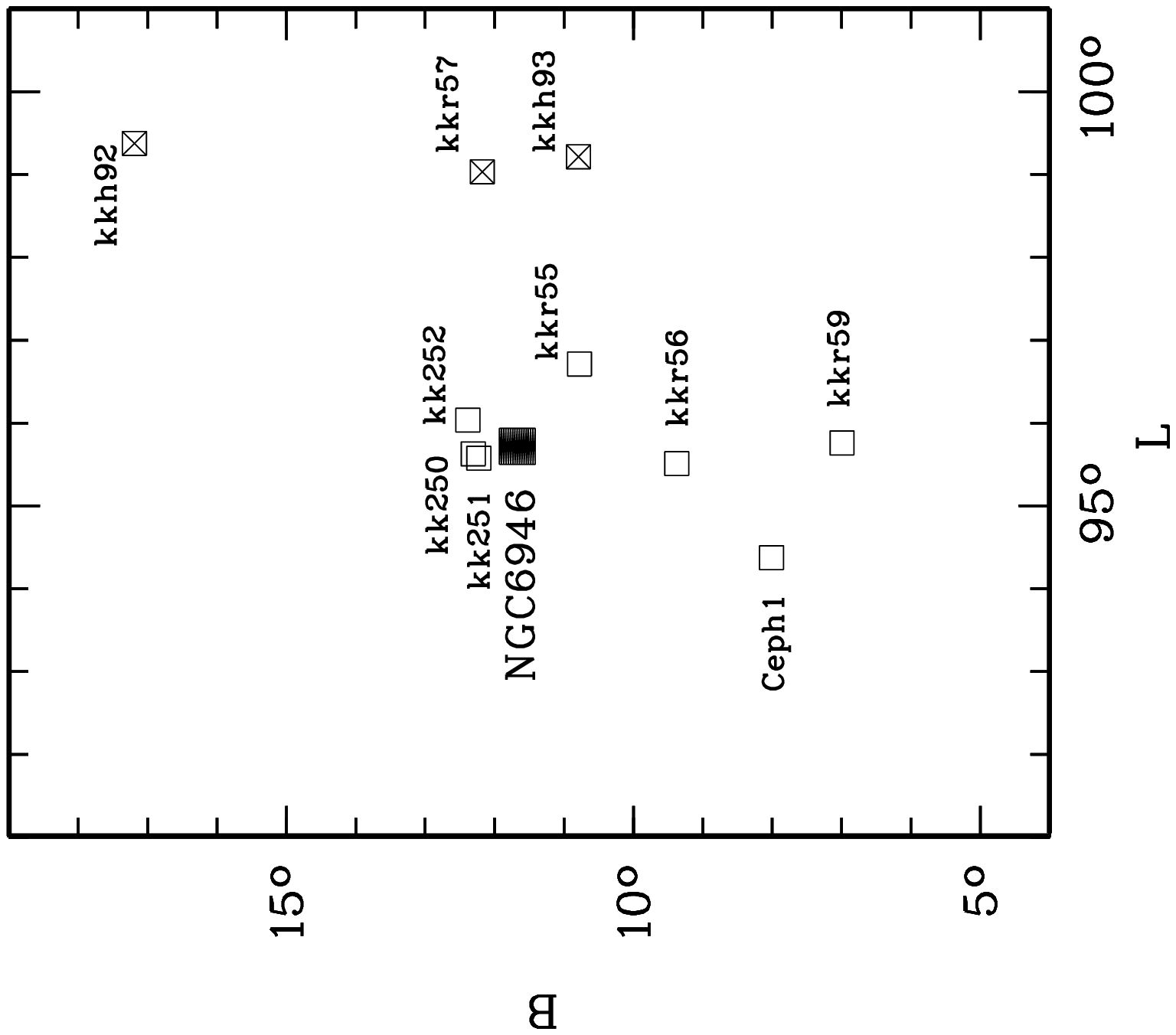}}\par
\vspace{10cm}
\caption{The distribution of LSB objects around NGC~6946 (full squares) in
       galactic coordinates. The open squares represent the members of the
       NGC~6946 group, the squares with cross correspond to other probable group members.}
\end{figure}
Probably the group contains also fainter members, but a search
for them would be a rather difficult observational task. Besides kk251,
where the estimate of $(B-V)_T$ is complicated by a superposition of bright
stars, the companions of NGC~6946 have colours typical for irregular
galaxies.

  Adopting the above identifications of HI emission in 
galaxies, we obtain hydrogen mass-to-luminosity ratios within
[0.18 -- 3.0] $M_{\sun}/L_{\sun}$, which is also typical for irregular galaxies.
The hydrogen mass is obtained from the HI flux, $F$ (in Jy$\cdot$km~s$^{-1}$), and distance
D (in Mpc), as

  $$   \log[M(HI)/M_{\sun}] = \log F + 2\log D + 5.37. $$

The mean square radial velocity of companions with respect to NGC~6946 is
95 km~s$^{-1}$, and their mean relative velocity is (+8$\pm$41) km~s$^{-1}$. Fig. 4 shows the
location of seven physical companions of NGC~6946 (open squares) in galactic
coordinates. Three other galaxies in the same area without radial velocities
and distance estimates are indicated with crosses. The most distant
group member, kkr59, is at 483 kpc from NGC~6946. This is
approximately the 
same size as that of the systems of companions we see around other near giant
spiral galaxies: M31, M81, and M101. Based on the mean linear separation
of dwarf galaxies in the NGC~6946 group, $<R>$ = 210 kpc, and their root mean square
radial velocity, 95 km~s$^{-1}$, we can estimate a virial mass of the group. In case
of randomly orientated closed orbits with a mean orbit eccentricity of
$e$ = 0.5 the virial (orbital) mass is 1.6$\cdot10^{12} M_{\sun}$. Since the integral
blue luminosity of the group is $2.9\cdot10^{10} L_{\sun}$, it yields a virial mass-to-
luminosity ratio $\sim56 M_{\sun}/L_{\sun}$, typical for small galaxy groups like
the Local Group.

\section{Concluding remarks}

  Being located at a low galactic latitude, $b = 12\degr$, in a zone of
substantial extinction, NGC~6946 looks at first glance like an isolated giant spiral galaxy
on the border of the Local Void (Tully, 1988). Recent searches
for low surface brightness galaxies on the POSS-II films revealed about
a dozen probable companions to NGC~6946. Subsequent HI observations
of the candidates and distance estimates from the brightest stars
showed that at least 7 dwarf irregular galaxies can be considered as
physical companions of NGC~6946. Therefore, a new nearby group of galaxies
has been discovered at a distance of (5.9$\pm$0.4) Mpc. The mean projected
diameter of the group is 420 kpc, its root mean square radial velocity is
95 km~s$^{-1}$, and the virial mass-to-luminosity ratio is $\sim56 M_{\sun}/L_{\sun}$,
typical for small galaxy groups. The two nearby giant face-on Sc galaxies:
M101 and NGC~6946, and their dwarfs are alike
with respect to their structure and kinematics. Being
gas-rich galaxies, the members of the NGC 6946 group need to be studied
in more detail with radio HI synthesis interferometry.

\acknowledgements{
We are grateful to A.Aparicio for providing us with CCD images of two objects
observed with the Nordic telescope.
This work has been partially supported by the DFG grant No 436 RUS 113/470/0.}

{}


\begin{thebibliography}{}

\bibitem{}Burton W.B., Braun R., Walterbos R.A., Hoopes C.G., 1999, AJ 117, 194
\bibitem{}de Vaucouleurs G., 1978, ApJ 224, 710
\bibitem{}Henning P.A., Kraan-Korteweg R.C., Rivers A.J., Loan A.J., Lahav O.,
   Burton W.B., 1998, AJ 115, 584
\bibitem{}Hopp U., 1998 (private information)
\bibitem{}Huchtmeier W.K., Karachentsev I.D., Karachentseva V.E., 1997, A\&A 322, 375
\bibitem{}Huchtmeier W.K., Karachentsev I.D., Karachentseva V.E., 2000, in Proc. IAU
   Colloq. No174, Turku, Finland
\bibitem{}Karachentsev I.D., Tikhonov N.A., 1994, A\&A 286, 718
\bibitem{}Karachentseva V.E., Karachentsev I.D., Huchtmeier W.K., 2000 (in preparation)
\bibitem{}Karachentseva V.E., Karachentsev I.D., 1998, A\&AS 127, 409
\bibitem{}Karachentseva V.E., Karachentsev I.D., Richter G.M., 1999, A\&AS 135, 221
\bibitem{}Landolt A., 1992, AJ 104, 340
\bibitem{}Makarova L.N., 1999, A\&AS 139, 491
\bibitem{}Sandage A., Tammann G.A., 1974, ApJ 191, 559
\bibitem{}Schlegel D.J., Finkbeiner D.P., Davis M., 1998, ApJ 500, 525
\bibitem{}Sharina M.E., Karachentsev I.D., Tikhonov N.A., 1997, Astr. Letters 23, 373
\bibitem{}Stetson P., Hesser J., Smecker-Hane T., 1998, PASP 110, 533
\bibitem{}Tully R.B., 1988, Nearby Galaxy Catalogue, Cambridge, Cambridge Univ. Press
\bibitem{}van Driel W., Kraan-Korteweg R.C., Binggeli B., Huchtmeier W.K., 1998,
   A\&AS 127, 397
\end{thebibliography}
\end{document}